\documentclass{article}[12pt]
\usepackage[utf8]{inputenc}
\usepackage{authblk}
\usepackage{graphicx}
\usepackage{amsmath}
\usepackage{natbib}
\usepackage{hyperref}
\usepackage{mhchem}
\usepackage{xcolor}
\usepackage{lineno}
\usepackage{here}
\usepackage[left=1in, right=1in, top=1in, bottom=1in]{geometry}
\title{Sensitive Probing of Exoplanetary Oxygen via Mid Infrared Collisional Absorption}

\author[1,2,3]{Thomas J. Fauchez}
\author[1,3]{Geronimo L. Villanueva}
\author[4,5,6,7,8]{Edward W. Schwieterman}
\author[9]{Martin Turbet}
\author[1,3,7]{Giada Arney}
\author[1,10]{Daria Pidhorodetska}
\author[1,3,7]{Ravi K. Kopparapu}
\author[1,3]{Avi Mandell}
\author[1,3,7]{Shawn D. Domagal-Goldman}

\affil[1]{\small NASA Goddard Space Flight Center, Greenbelt, Maryland, USA}
\affil[2]{\small Goddard Earth Sciences Technology and Research (GESTAR), Universities Space Research Association, Columbia, Maryland, USA}
\affil[3]{\small GSFC Sellers Exoplanet Environments Collaboration}
\affil[4]{\small Department of Earth and Planetary Sciences, University of California, Riverside, California, USA}
\affil[5]{\small NASA Postdoctoral Program, Universities Space Research Association, Columbia, Maryland, USA}
\affil[6]{\small NASA Astrobiology Institute, Alternative Earths Team, Riverside, CA, USA}
\affil[7]{\small Nexus for Exoplanet System Science (NExSS) Virtual Planetary Laboratory, Seattle, WA, USA} 
\affil[8]{\small Blue Marble Space Institute of Science, Seattle, Washington, USA}
\affil[9]{\small Observatoire  Astronomique  de  l’Universit\'e  de  Gen\`eve,  Universit\'e  de  Gen\`eve,  Chemin  des Maillettes 51, 1290 Versoix, Switzerland.}
\affil[10]{\small University of Maryland Baltimore County/CRESST II \\
1000 Hilltop Cir. Baltimore, MD 21250, USA}

\date{Accepted in Nature Astronomy on 10/01/2019 \\ Published in Nature astronomy on 06/01/2020}

\begin{document}

\maketitle
\begin{abstract}
\textbf{The collision-induced fundamental vibration-rotation band at $6.4\ \mu m$  is the most significant absorption feature from O$_2$ in the infrared \citep{Timofeyev1978,Rinsland1982,Rinsland1989}, yet it has not been previously incorporated into exoplanet spectral analyses for several reasons.
 Either CIAs were not included or incomplete/obsolete CIA databases were used.  Also, the current version of HITRAN does not include CIAs at  $6.4\ \mu m$ with other collision partners (O$_2$-X). We include O$_2$-X CIA features in our transmission spectroscopy simulations by parameterizing the $6.4\ \mu m$ O$_2$-N$_2$ CIA  based on \cite{Rinsland1989} and the O$_2$-CO$_2$ CIA based on \cite{Baranov2004}. Here we report  that  the O$_2$-X  CIA may be the most detectable O$_2$ feature for transit observations. For a potential TRAPPIST-1e analogue system within 5~pc of the Sun, it could be the only O$_2$ detectable signature with JWST (using MIRI LRS) for a modern Earth-like cloudy atmosphere with biological quantities of O$_2$.  Also, we show that the $6.4\ \mu m$ O$_2$--X CIA would be prominent for O$_2$-rich desiccated atmospheres \citep{LugerBarnes2015} and could be detectable with JWST in just a few transits. For systems beyond 5~pc, this feature could therefore be a powerful discriminator of uninhabited planets with non-biological "false positive" O$_2$ in their atmospheres - as they would only be detectable at those higher O$_2$ pressures.}

\end{abstract}
\section*{Main}

 We study the strength of the O$_2$-X CIA spectral signatures in exoplanets by computing synthetic spectra for various Earth-like atmospheres with the Planetary Spectrum Generator  (PSG \cite{Villanueva2018}). The atmospheres are created with the LMD-G \citep{Wordsworth2011} general circulation model (GCM) coupled with the Atmos \citep{Arney2016} photochemical model (see Methods for details).  We focus in particular on planets around M dwarfs such as TRAPPIST-1e. In fact, For modern Earth atmospheric conditions, the $6.4\ \mu m$ region is overlapped by a wide H$_2$O absorption band.  However, for a modern Earth-like atmosphere on a tidally locked planet in the HZ of an M dwarf, the terminator region is predicted to be fairly dry (see Supplementary Figure 2). Also, water is mostly confined in a small portion of the atmosphere near the surface and which is under the refraction limit and hidden by clouds (as on Earth, where the troposphere is wet and the stratosphere is dry).  Near TOA, H$_2$O is highly photodissociated (Supplementary Figure 2). The H$_2$O signature in the transmission spectra of a habitable planet is therefore expected to be very weak \citep{Lincowski2018,Lustig-Yaeger2019}. While some trace gases such as NO$_2$ and N$_2$O also produce opacity in this spectral region, their concentrations are predicted to be orders of magnitude lower than those that would generate confounding impacts on the simulated spectra. \\

\begin{figure}[h!]
\centering
\resizebox{14cm}{!}{\includegraphics{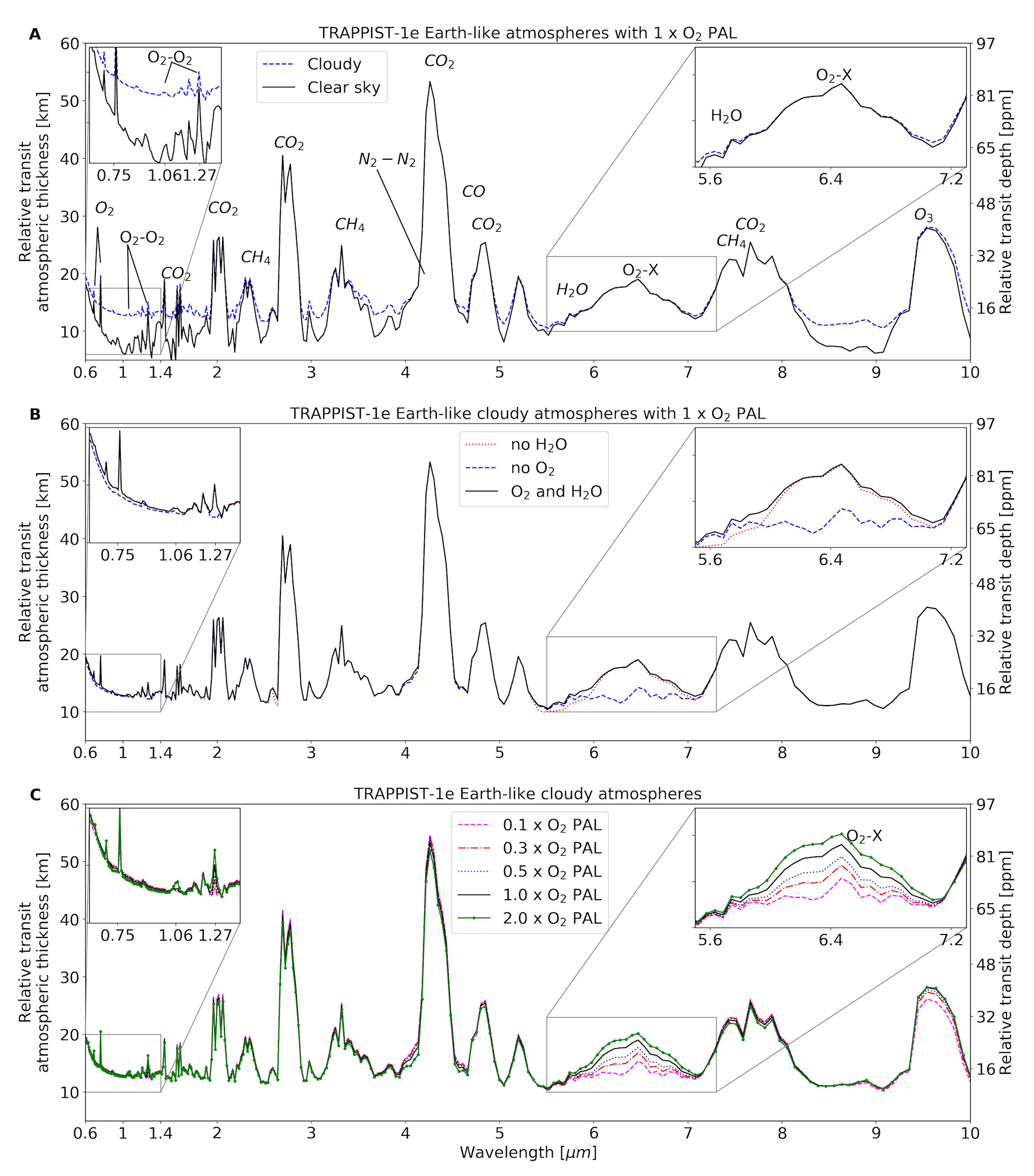}}
\caption{Shown are Earth-like transmission spectra of TRAPPIST-1e. Panel A: the impact of cloud coverage on the atmosphere's spectral features.  Panel B: a comparison of the strengths of the O$_2$-X CIA feature and the H$_2$O absorption band around $6.4\ \mu m$ for a cloudy atmosphere.  Panel C: the strength of O$_2$ monomer absorptions and CIA features as a function of the amount of O$_2$ in the atmosphere relative to PAL for a spectrum with clouds included. The O$_2$-X CIA could be the strongest O$_2$ feature across the VIS/NIR/MIR spectrum. }
\label{fig:proto}
\end{figure}

The TRAPPIST-1 system \citep{Gillon2017}, consisting of seven Earth-sized planets orbiting an ultra-cool dwarf star, will be a favorite target for atmospheric characterization with JWST due to its relatively close proximity to the Earth and the depth and frequency of its planetary transits. Therefore, we use TRAPPIST-1e as a case study for our simulated spectra. We employed the LMD-G \citep{Wordsworth2011} general circulation model (GCM) and the Atmos photochemical model \citep{Arney2016} to simulate TRAPPIST-1e with boundary conditions similar to modern Earth \citep{Lincowski2018}. \\
Figure \ref{fig:proto} shows TRAPPIST-1e transmission spectra from 0.6 to 10 $\mu m$ for various Earth-like atmospheres simulated with the Planetary Spectrum Generator (PSG, \cite{Villanueva2018}). The top panel shows the impact of cloud coverage on spectral features: clouds diminish the strength of all absorption features, but impact the strength of the O$_2$-X feature much less strongly than they impact shorter wavelength O$_2$ features like the O$_2$ A--band or the 1.06 and 1.27 $\mu m$ O$_2$ CIA  used in \citet{Misra2014} (who considered only clear-sky atmospheres). This is because water cloud opacity is stronger at short wavelengths.  The middle panel compares the strength of the O$_2$-X CIA band to the overlapping H$_2$O absorption band near $6.4\ \mu m$ for a cloudy atmosphere.  O$_2$-X CIA strongly dominates the absorption in this wavelength range. The bottom panel shows how the strengths of O$_2$ monomer and CIA absorption features scale as a function of the O$_2$ atmospheric abundance for O$_2$ levels ranging from 0.1 times the present atmospheric level of O$_2$ (PAL) to 2 times PAL. 
Our results show that the $6.4\ \mu m$ CIA feature appears to be about three times stronger than the 1.27 $\mu m$  O$_2$ CIA feature and is therefore the strongest O$_2$ signature across the VIS/NIR/MIR spectrum. \\

\begin{figure}[h!]
\centering
\resizebox{15cm}{!}{\includegraphics{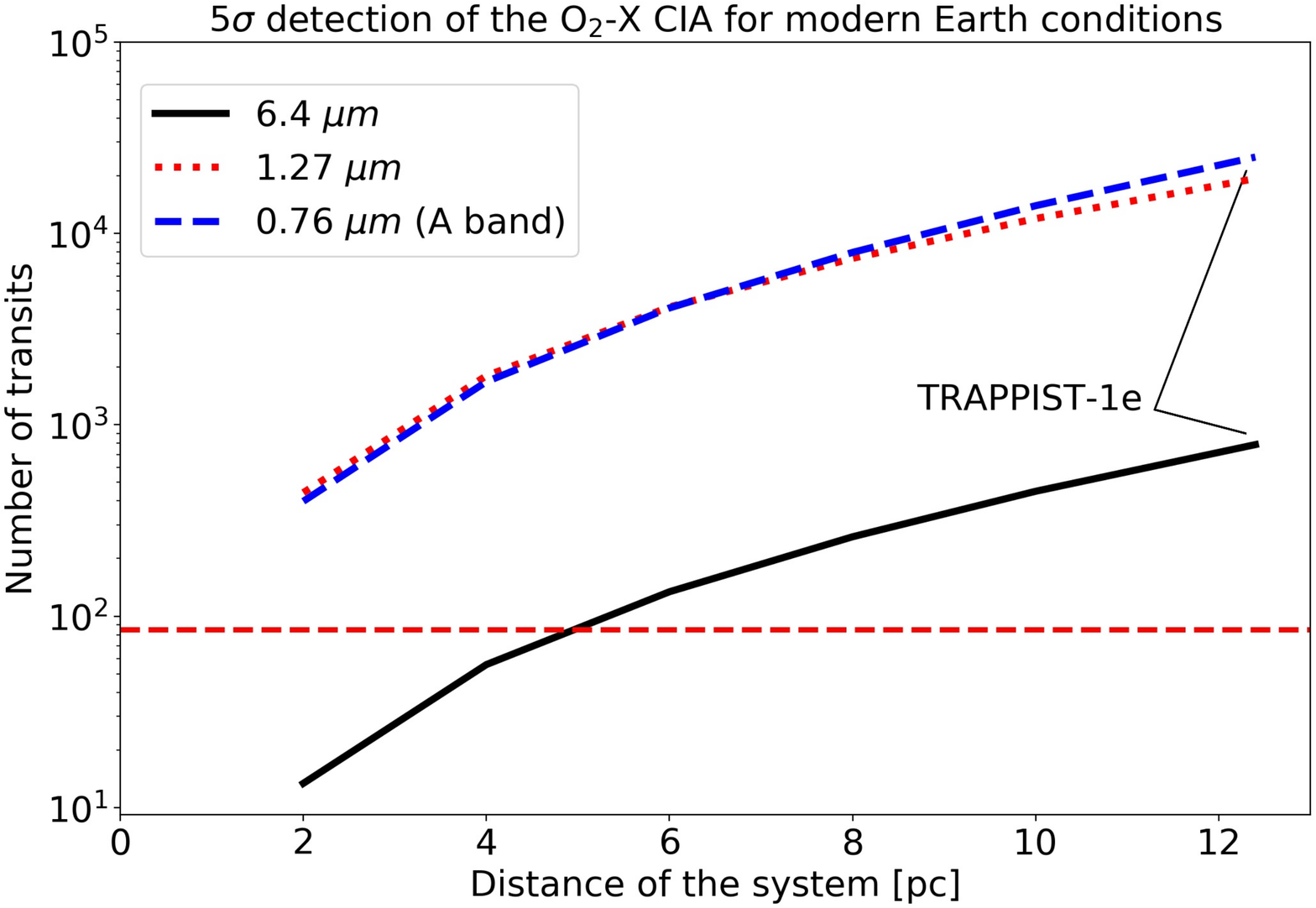}}
\caption{Number of TRAPPIST-1e transits needed for a $5\ \sigma$ detection of the O$_2$ A--band (R=100), the O$_2$-O$_2$ CIA at 1.27 $\mu m$ (R=100) and the O$_2$-X CIA at 6.4 $\mu m$ (R=30) with JWST for the TRAPPIST-1 system moved from its distance to Sun (12.1~pc) down to 2~pc. The atmosphere is composed of N$_2$, 10,000~ppm of CO$_2$, 10~ppm of CH$_4$, 21$\%$ of O$_2$ with a surface pressure of 1~bar. Resolving power (R) has been optimized for each band to maximize the SNR. The horizontal dashed red line corresponds to the number of times TRAPPIST-1e will be observable transiting in front of TRAPPIST-1 during JWST 5.5 years life time (85 transits). The 6.4 $\mu m$  O$_2$-X CIA requires much less transits than the O$_2$ A--band and the 1.27 $\mu m$ O$_2$-O$_2$ CIA and can be detectable at $5\ \sigma$ for a  TRAPPIST-1/TRAPPIST-1e analogue system closer than 5~pc.}
\label{fig:modernTransits}
\end{figure}

Figure \ref{fig:modernTransits} shows the number of TRAPPIST-1e transits needed to detect the O$_2$ A--band, the O$_2$-O$_2$ CIA feature at 1.27 $\mu m$ and the O$_2$-X CIA feature at 6.4 $\mu m$ features at a $5\ \sigma$ confidence level with JWST for a modern Earth-like cloudy atmosphere on TRAPPIST-1e orbiting a TRAPPIST-1-like star at distances from Earth ranging from TRAPPIST-1's true distance (12.1~pc) down to 2~pc. We can see that the 6.4 $\mu m$ O$_2$-X CIA feature requires an order of magnitude fewer transits than the two other O$_2$ features because of the stronger intrinsic O$_2$-X CIA absorption at 6.4 $\mu m$ and because cloud opacity is stronger at shorter VIS/NIR wavelengths. The horizontal dashed red line represents the 85 transits that will occur for TRAPPIST-1e during the 5.5 year nominal lifetime  of JWST, thus sets up an upper limit on the number of transits observable. Because TRAPPIST-1e orbits a very small M8 star, it offers one of the best SNR a habitable planet can have and therefore represents a best-case scenario in terms of detectability. However, even in this context, none of the O$_2$ features are detectable at $5\ \sigma$ at the distance of the TRAPPIST-1 system. However, the 6.4 $\mu m$ O$_2$-X CIA feature could be detectable at $5\ \sigma$ for an analogue system at star-Earth distance closer than 5~pc.  Therefore, this simulation shows that the 6.4 $\mu m$ O$_2$-X CIA could be the only oxygen feature detectable with JWST for a cloudy modern Earth-like atmosphere for nearby hypothetical TRAPPIST-1 analogue systems. \\

The O$_2$-X feature for oxygen could also potentially be used to detect non-habitable conditions, such as a desiccated atmosphere rich in bars of abiotic O$_2$ generated from massive ocean loss  \citep{Wordsworth2014,LugerBarnes2015,Schwieterman2016,Meadows2017,Meadows2018,Lustig-Yaeger2019}. \cite{Lincowski2018} have shown that for an assumed original water content of 20 Earth oceans (by mass), the TRAPPIST-1e, f and g planets may have lost between 3 to 6 Earth oceans resulting in atmospheres with 22 and 5,000 bars of O$_2$. \\

\begin{figure}[h!]
\centering
\resizebox{15cm}{!}{\includegraphics{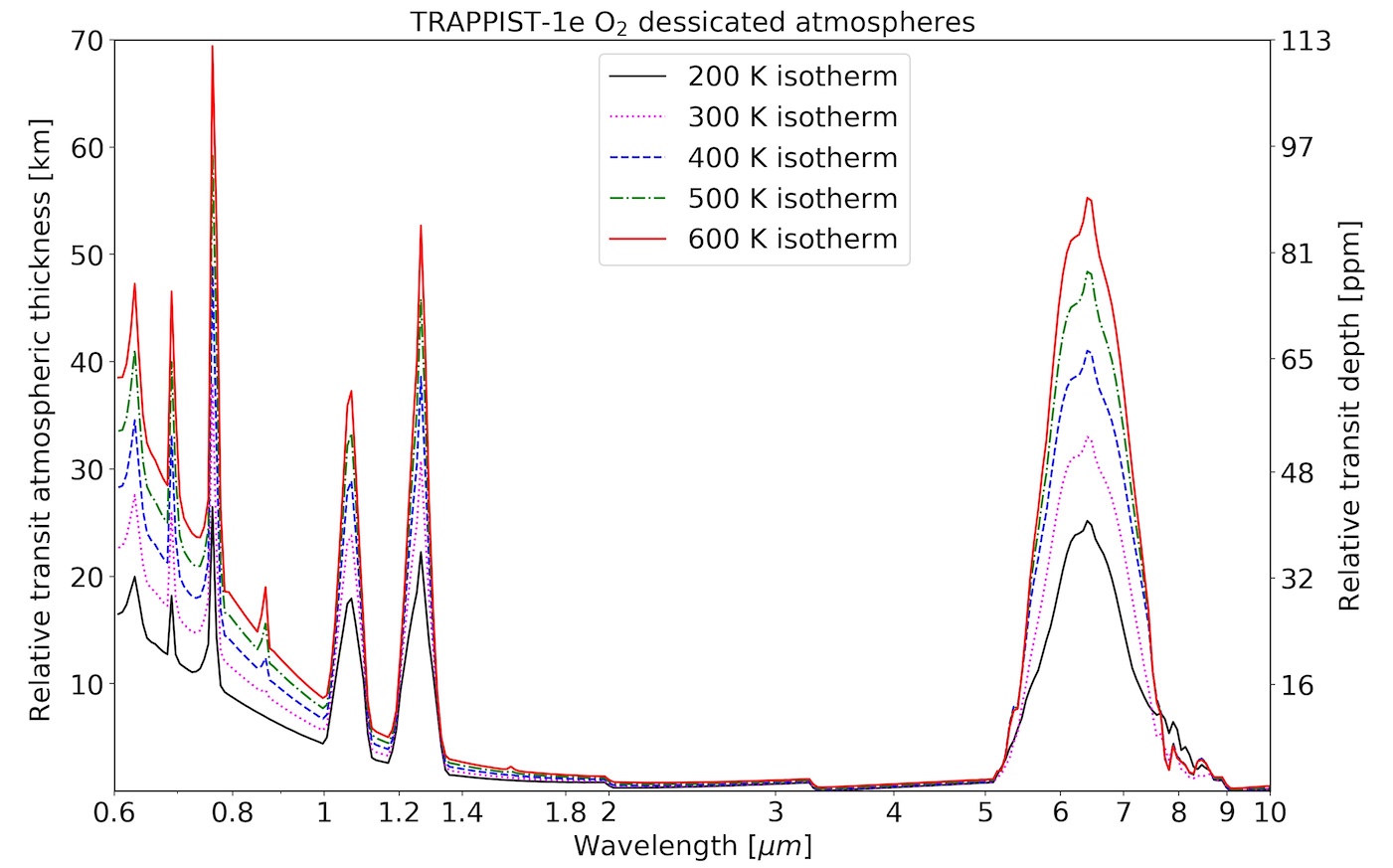}}
\caption{Transmission spectra for a 1~bar O$_2$ desiccated atmosphere on TRAPPIST-1e assuming various isothermal profiles. Depending on the temperature, the $6.4\ \mu m$ O$_2$-O$_2$ CIA feature can reach between 40 to 90~ppm, which is comparable or larger to the O$_2$ A--band and 1.06 and 1.27 $\mu m$  O$_2$ CIA features. No photochemistry is considered here so O$_3$ is missing in the spectra. Note that the increase of the relative transit atmospheric thickness and relative transit depth is due to the increase of the scale height with temperature.}
\label{fig:dessic}
\end{figure}

Figure \ref{fig:dessic} shows transit spectra for TRAPPIST-1e assuming conservative 1~bar O$_2$-only desiccated and isothermal atmospheres ranging from 200 to 600 K. Relative transit depth (ppm, left Y-axis) is the transit depth produced by the atmosphere itself, which can be converted into relative transit atmospheric thickness (km, right Y-axis). These isothermal profiles allow us to test the sensitivity of oxygen spectral features on atmospheric temperature. The atmospheric scale height increases with temperature, and the largest features are seen for the highest temperatures. Note that O$_2$-O$_2$ CIA opacities in HITRAN are only  provided in the 193~K--353~K temperature range. Therefore, for the isothermal profiles beyond 353~K we used the 353-K CIA coefficients. 
We can see  that the  $6.4\ \mu m$ O$_2$-O$_2$  CIA feature is broad ($\sim \ 3\ \mu m$) and strong (40 to 90~ppm). The $1.27\ \mu m$ O$_2$--O$_2$ CIA feature reaches a similar relative transit depth but is comparatively narrower (widths of $\sim0.2\ \mu m$). In addition,   the continuum level for the shorter wavelengths is raised by Rayleigh scattering slope, reducing the NIR CIA relative transit depths to 50 to 80~ppm, respectively. Similarly, the O$_2$ A--band reaches very high transit depths (up to 110~ppm) but on a high continuum, which reduces its relative strength down to 95~ppm. 
 The larger width of the O$_2$-X CIA feature at $6.4\ \mu m$ allows us to bin down the data to a lower resolving power, improving the SNR and therefore compensating for a higher noise floor in the MIRI LRS range. Supplementary Table 1 presents the relative transit depth, 1 transit SNR and number of transits for 3 and 5 $\sigma$ detections for TRAPPIST-1e assuming 1 and  22~bar desiccated atmosphere on TRAPPIST-1e and Supplementary Figure 3 is similar to Fig. \ref{fig:modernTransits} but for the 22 bar O$_2$ desiccated and isothermal atmospheres. \\

\newpage

 Interpreting an O$_2$ detection via the O$_2$-O$_2$ CIA band at $6.4\ \mu m$ will be strengthened by constraining the concentration of O$_2$, and placing its presence in a broader atmospheric context. For HZ planets with a planet/star contrast comparable to TRAPPIST-1e and within 5~pc from the Sun, next-generation MIR observatories could detect O$_2$ in concentrations similar to modern Earth using the $6.4\ \mu m$ O$_2$-X feature. In combination with detections of other MIR features from CH$_4$, H$_2$O, or N$_2$O, this would represent a strong biosignature with no known non-biological explanations \citep{DesMarais2002}. Note that there are 50 red dwarfs within 5 pc from the Sun (http://www.recons.org/TOP100.posted.htm).\\
 For systems farther than about 5~pc and/or HZ planet orbiting earlier M dwarfs, JWST or future MIR observatories may be able to detect the $6.4\ \mu m$ O$_2$-X feature only for O$_2$ concentrations orders of magnitude higher than those on modern-day Earth that would be indicative of a desiccated, O$_2$-rich, uninhabitable planet. Detection of this feature for planets within the habitable zone \citep{Kasting1993, Kopparapu2013,Kopparapu2014} will test the hypothesis that the high luminosity pre-main sequence phase M dwarfs endure can render even current HZ planets uninhabitable \citep{LugerBarnes2015}.  Finally, detection of this feature would answer the question of whether planets around M dwarfs can sustain an atmosphere.\\

\section*{Methods}
\paragraph{Parameters for TRAPPIST-1e}
 In this study, TRAPPIST-1e planet's parameters have been set up from \citep{Gillon2017, Grimm2018}. The TRAPPIST-1 spectrum of \cite{Lincowski2018} has been used for our photochemical simulations with the Atmos model.

\paragraph{Monomer and CIA pressure sensitivity}
Monomer and CIA optical depths can be expressed by the following equations \citep{Misra2014}:
\begin{equation}
    d\tau_{mono}=\sigma\rho dl = \sigma P/T dl
\end{equation}
\begin{equation}
    d\tau_{CIA}=k\rho^2 dl = k (P/T)^2 dl
\end{equation}

with $ d\tau_{mono}$ and $d\tau_{CIA}$ representing the monomer and CIA differential optical depths, respectively; $\sigma$ and $k$ are the monomer and CIA cross sections, respectively; $\rho$ is the number density of the gas; P is the pressure; T is the temperature; and $dl$ is the path length. 
$d\tau_{mono}$ is proportional to P and $d\tau_{CIA}$ to $P^2$, and this difference of sensitivity may be used to estimate the atmospheric pressure \citep{Misra2014}.

\paragraph{The atmospheric modeling}
We use the Atmos \citep{Arney2016} photochemical model to self-consistently simulate Earth-like atmospheres with a variety of O$_2$ partial pressures on TRAPPIST-1e. The terminator temperature, gas mixing ratio, vapor and condensed water (liquid and ice) profiles have been provided from the LMD-G \citep{Wordsworth2011}  global climate model (GCM) simulations of a 1~bar TRAPPIST-1e modern Earth atmosphere. In Atmos, some of the N$_2$ has been swapped for O$_2$ to obtain various O$_2$ PAL as shown in Fig. \ref{fig:proto}, both gases having no greenhouse effect except through pressure broadening or CIA, and 10,000~ppm of CO$_2$ and 10~ppm of CH$_4$ have been assumed. Due to the terminator atmospheric profiles varying with latitude, Atmos was used to calculate profiles  for 98 latitude points, determined by the LMD-G latitude resolution.

\paragraph{Transmission spectra simulations}
The planetary spectrum generator (PSG, \cite{Villanueva2018}) has been used to simulate JWST transmission spectra. PSG is an online radiative-transfer code that is able to compute planetary spectra (atmospheres and surfaces) for a wide range of wavelengths (UV/Vis/near-IR/IR/far-IR/THz/sub-mm/Radio) from any observatory, orbiter or lander and also includes a noise calculator. 
To compute the noise, PSG  takes into account  the noise introduced by the source itself (N$_{source}$), the background noise (N$_{back}$) following a Poisson distribution with fluctuations depending on $\sqrt{N}$ with  $N$  the mean number of photons received \citep{Zmuidzinas2003}, the noise of the detector (N$_{D}$) and the noise introduced by the telescope (N$_{optics}$). The total noise being then $N_{total}=\sqrt{N_{source}+N_{back}+N_D+N_{optics}}$.  This represents therefore a photon limited situation where N$_{source}$ will largely dominate $N_{total}$.\\
For the Earth-like atmospheres, spectra were obtained for each of the 98 Atmos photochemical simulations and an average spectrum was computed. For the 1 and 22~bar O$_2$ desiccated atmospheres, isothermal profiles from 200 to 600~K were set up with 100\% of O$_2$, ignoring photochemistry. To calculate the SNR and the number of transits needed for 3 and 5$\sigma$ detection  the resolving power has been optimized by adjusting the binning for each O$_2$ feature to maximize its SNR. SNR is calculated using the highest value in the band minus the nearest continuum value (this value therefore differ between the VIS (Rayleigh slope), NIR and MIR). The number of transits needed to achieve a X$\sigma$ detection is computed as the following equation:
\begin{equation}\label{eq:SNRi}
    N_{transits}^{X\sigma}= N_i*(X/SNR_i)^2
\end{equation}

with $X\sigma$ the confident level of value $X$, N$_i$ is the initial number of transit at which SNR$_i$ is computed. If SNR$_i$ is estimated from 1 transit then $N_i=1$ and the Eq. \ref{eq:SNRi} could be simplified as:
\begin{equation}
    N_{transits}^{X\sigma}= (X/SNR_i)^2
\end{equation}

\paragraph{O$_2$-X collision-induced absorption at $6.4\ \mu m$.}
This feature is associated with the fundamental band of O$_2$, and O$_2$ collisions with other partners (e.g. N$_2$, CO$_2$) can produce additional absorption at these wavelengths. This collision with other gases can be generally written as O$_2$-X, where "X" refers to the collision partner.  Laboratory measurements \citep{Timofeyev1978,Thibault1997} and atmospheric analysis using Sun occultations \citep{Rinsland1989} have revealed that nitrogen, the major constituent of modern Earth's atmosphere at $78\%$ in volume, produces an O$_2$--N$_2$ absorption feature of a similar intensity as O$_2$--O$_2$ in the $6.4\ \mu m$ region. Carbon dioxide (CO$_2$) can also produce an O$_2$--CO$_2$ feature at these wavelengths, though this feature is weak for modern Earth-like CO$_2$ atmospheric abundances (approx. 400 ppm) but can be strong for exoplanets with CO$_2$ rich atmospheres  \citep{Baranov2004}.  O$_2$-X CIA can also be produced with H$_2$O \citep{Hopfner2012} as the collision partner due to the large electric dipole moment of H$_2$O, but no laboratory measurements exist for this feature. \\

\paragraph{Parameterization of the 6.4 $\mathbf{\mu m}$ feature}
While the $6.4\ \mu m$ region is known as the fundamental vibration-rotation band of O$_2$, only the O$_2$-O$_2$ CIA band is included in HITRAN \citep{Gordon2017}. Knowing that Earth's atmosphere is mostly composed of N$_2$ and that the O$_2$-N$_2$ CIA have been shown to produce similar absorption to O$_2$-O$_2$ \citep{Timofeyev1978,Rinsland1989,Thibault1997}, it is important to include it in our simulations. We have parameterized the O$_2$-N$_2$ CIA in PSG assuming the same absorption efficiency as O$_2$-O$_2$ CIA \citep{Rinsland1989} (see Supplementary Figure 1). For O$_2$-CO$_2$ CIA at $6.4\ \mu m$, we used experimental data of \cite{Baranov2004} to include that feature in PSG. 

\bibliographystyle{aasjournal.bst}
\bibliography{Bibref.bib}

\begin{thebibliography}{}
\expandafter\ifx\csname natexlab\endcsname\relax\def\natexlab#1{#1}\fi
\providecommand{\url}[1]{\href{#1}{#1}}

\bibitem[{{Arney} {et~al.}(2016){Arney}, {Domagal-Goldman}, {Meadows}, {Wolf},
  {Schwieterman}, {Charnay}, {Claire}, {H{\'e}brard}, \& {Trainer}}]{Arney2016}
{Arney}, G., {Domagal-Goldman}, S.~D., {Meadows}, V.~S., {et~al.} 2016,
  Astrobiology, 16, 873

\bibitem[{Baranov {et~al.}(2004)Baranov, Lafferty, \& Fraser}]{Baranov2004}
Baranov, Y.~I., Lafferty, W., \& Fraser, G. 2004, J MOL SPECTROSC, 228, 432 ,
  special Issue Dedicated to Dr. Jon T. Hougen on the Occasion of His 68th
  Birthday.
\newblock
  \url{http://www.sciencedirect.com/science/article/pii/S0022285204001390}

\bibitem[{{Des Marais} {et~al.}(2002){Des Marais}, {Harwit}, {Jucks},
  {Kasting}, {Lin}, {Lunine}, {Schneider}, {Seager}, {Traub}, \&
  {Woolf}}]{DesMarais2002}
{Des Marais}, D.~J., {Harwit}, M.~O., {Jucks}, K.~W., {et~al.} 2002,
  Astrobiology, 2, 153

\bibitem[{Fauchez {et~al.}(2017)Fauchez, Rossi, \& Stam}]{Fauchez2017}
Fauchez, T., Rossi, L., \& Stam, D.~M. 2017, ASTROPHYS J, 842, 41

\bibitem[{Gillon {et~al.}(2017)Gillon, Triaud, Demory, Jehin, Agol, Deck,
  Lederer, de~Wit, Burdanov, Ingalls, Bolmont, Leconte, Raymond, Selsis,
  Turbet, Barkaoui, Burgasser, Burleigh, Carey, Chaushev, Copperwheat, Delrez,
  Fernandes, Holdsworth, Kotze, Van~Grootel, Almleaky, Benkhaldoun, Magain, \&
  Queloz}]{Gillon2017}
Gillon, M., Triaud, A. H. M.~J., Demory, B.-O., {et~al.} 2017, Nature, 542,
  456–460.
\newblock \url{https://doi.org/10.1038/nature21360}

\bibitem[{Gordon {et~al.}(2017)Gordon, Rothman, Hill, Kochanov, Tan, Bernath,
  Birk, Boudon, Campargue, Chance, Drouin, Flaud, Gamache, Hodges, Jacquemart,
  Perevalov, Perrin, Shine, Smith, Tennyson, Toon, Tran, Tyuterev, Barbe,
  Csaszar, Devi, Furtenbacher, Harrison, Hartmann, Jolly, Johnson, Karman,
  Kleiner, Kyuberis, Loos, Lyulin, Massie, Mikhailenko, Moazzen-Ahmadi, Muller,
  Naumenko, Nikitin, Polyansky, Rey, Rotger, Sharpe, Sung, Starikova, Tashkun,
  Auwera, Wagner, Wilzewski, Wcisto, Yu, \& Zak}]{Gordon2017}
Gordon, I., Rothman, L., Hill, C., {et~al.} 2017, J QUANT SPECTROSC RA, 203, 3
  , hITRAN2016 Special Issue.
\newblock
  \url{http://www.sciencedirect.com/science/article/pii/S0022407317301073}

\bibitem[{{Grimm} {et~al.}(2018){Grimm}, {Demory}, {Gillon}, {Dorn}, {Agol},
  {Burdanov}, {Delrez}, {Sestovic}, {Triaud}, {Turbet}, {Bolmont}, {Caldas},
  {Wit}, {Jehin}, {Leconte}, {Raymond}, {Grootel}, {Burgasser}, {Carey},
  {Fabrycky}, {Heng}, {Hernandez}, {Ingalls}, {Lederer}, {Selsis}, \&
  {Queloz}}]{Grimm2018}
{Grimm}, S.~L., {Demory}, B.-O., {Gillon}, M., {et~al.} 2018, ASTRON ASTROPHYS,
  613, A68

\bibitem[{Hopfner {et~al.}(2012)Hopfner, Milz, Buehler, Orphal, \&
  Stiller}]{Hopfner2012}
Hopfner, M., Milz, M., Buehler, S., Orphal, J., \& Stiller, G. 2012, GEOPHYS
  RES LETT, 39,
  https://agupubs.onlinelibrary.wiley.com/doi/pdf/10.1029/2012GL051409

\bibitem[{Kasting {et~al.}(1993)Kasting, Whitmire, \& Reynolds}]{Kasting1993}
Kasting, J.~F., Whitmire, D.~P., \& Reynolds, R.~T. 1993, Icarus, 101, 108 .
\newblock
  \url{http://www.sciencedirect.com/science/article/pii/S0019103583710109}

\bibitem[{Kopparapu {et~al.}(2014)Kopparapu, Ramirez, SchottelKotte, Kasting,
  Domagal-Goldman, \& Eymet}]{Kopparapu2014}
Kopparapu, R.~K., Ramirez, R.~M., SchottelKotte, J., {et~al.} 2014, ASTROPHYS
  J, 787, L29.
\newblock \url{https://doi.org/10.1088%2F2041-8205%2F787%2F2%2Fl29}

\bibitem[{Kopparapu {et~al.}(2013)Kopparapu, Ramirez, Kasting, Eymet, Robinson,
  Mahadevan, Terrien, Domagal-Goldman, Meadows, \& Deshpande}]{Kopparapu2013}
Kopparapu, R.~K., Ramirez, R., Kasting, J.~F., {et~al.} 2013, ASTROPHYS J, 765,
  131.
\newblock \url{http://stacks.iop.org/0004-637X/765/i=2/a=131}

\bibitem[{{Lincowski} {et~al.}(2018){Lincowski}, {Meadows}, {Crisp},
  {Robinson}, {Luger}, {Lustig-Yaeger}, \& {Arney}}]{Lincowski2018}
{Lincowski}, A.~P., {Meadows}, V.~S., {Crisp}, D., {et~al.} 2018, ASTROPHYS J,
  867, 76

\bibitem[{{Luger} \& {Barnes}(2015)}]{LugerBarnes2015}
{Luger}, R., \& {Barnes}, R. 2015, Astrobiology, 15, 119

\bibitem[{Lustig-Yaeger {et~al.}(2019)Lustig-Yaeger, Meadows, \&
  Lincowski}]{Lustig-Yaeger2019}
Lustig-Yaeger, J., Meadows, V.~S., \& Lincowski, A.~P. 2019, ASTRON J,
  \textbf{158}, 27.
\newblock \url{https://doi.org/10.3847%2F1538-3881%2Fab21e0}

\bibitem[{Meadows(2017)}]{Meadows2017}
Meadows, V.~S. 2017, Astrobiology, 17, 1022, pMID: 28443722.
\newblock \url{https://doi.org/10.1089/ast.2016.1578}

\bibitem[{Meadows {et~al.}(2018)Meadows, Reinhard, Arney, Parenteau,
  Schwieterman, Domagal-Goldman, Lincowski, Stapelfeldt, Rauer, DasSarma,
  Hegde, Narita, Deitrick, Lustig-Yaeger, Lyons, Siegler, \&
  Grenfell}]{Meadows2018}
Meadows, V.~S., Reinhard, C.~T., Arney, G.~N., {et~al.} 2018, Astrobiology, 18,
  630, pMID: 29746149.
\newblock \url{https://doi.org/10.1089/ast.2017.1727}

\bibitem[{Misra {et~al.}(2014)Misra, Meadows, Claire, \& Crisp}]{Misra2014}
Misra, A., Meadows, V., Claire, M., \& Crisp, D. 2014, Astrobiology, 14, 67

\bibitem[{{Pall{\'e}} {et~al.}(2009){Pall{\'e}}, {Zapatero Osorio}, {Barrena},
  {Monta{\~n}{\'e}s-Rodr{\'{\i}}guez}, \& {Mart{\'{\i}}n}}]{Palle2009}
{Pall{\'e}}, E., {Zapatero Osorio}, M.~R., {Barrena}, R.,
  {Monta{\~n}{\'e}s-Rodr{\'{\i}}guez}, P., \& {Mart{\'{\i}}n}, E.~L. 2009,
  Nature, 459, 814

\bibitem[{Rinsland {et~al.}(1982)Rinsland, Smith, Seals~Jr., Goldman, Murcray,
  Murcray, Larsen, \& Rarig}]{Rinsland1982}
Rinsland, C.~P., Smith, M. A.~H., Seals~Jr., R.~K., {et~al.} 1982, J GEOPHYS
  RES-OCEANS, 87, 3119

\bibitem[{Rinsland {et~al.}(1989)Rinsland, Zander, Namkung, Farmer, \&
  Norton}]{Rinsland1989}
Rinsland, C.~P., Zander, R., Namkung, J.~S., Farmer, C.~B., \& Norton, R.~H.
  1989, J GEOPHYS RES-ATMOS, 94, 16303

\bibitem[{{Schwieterman} {et~al.}(2016){Schwieterman}, {Meadows},
  {Domagal-Goldman}, {Deming}, {Arney}, {Luger}, {Harman}, {Misra}, \&
  {Barnes}}]{Schwieterman2016}
{Schwieterman}, E.~W., {Meadows}, V.~S., {Domagal-Goldman}, S.~D., {et~al.}
  2016, ASTROPHYS J, 819, L13

\bibitem[{Snellen {et~al.}(2013)Snellen, de~Kok, le~Poole, Brogi, \&
  Birkby}]{Snellen2013}
Snellen, I. A.~G., de~Kok, R.~J., le~Poole, R., Brogi, M., \& Birkby, J. 2013,
  ASTROPHYS J, 764, 182.
\newblock \url{https://doi.org/10.1088%2F0004-637x%2F764%2F2%2F182}

\bibitem[{Thibault {et~al.}(1997)Thibault, Menoux, Doucen, Rosenmann, Hartmann,
  \& Boulet}]{Thibault1997}
Thibault, F., Menoux, V., Doucen, R.~L., {et~al.} 1997, APPL OPTIC, 36, 563.
\newblock \url{http://ao.osa.org/abstract.cfm?URI=ao-36-3-563}

\bibitem[{Timofeyev \& Tonkov(1978)}]{Timofeyev1978}
Timofeyev, Y., \& Tonkov, M. 1978, lzv. Acad. Sci. USSR Atmos. Ocean. Phys.,
  Engl. Transl., 14, 614

\bibitem[{{Villanueva} {et~al.}(2018){Villanueva}, {Smith}, {Protopapa},
  {Faggi}, \& {Mandell}}]{Villanueva2018}
{Villanueva}, G.~L., {Smith}, M.~D., {Protopapa}, S., {Faggi}, S., \&
  {Mandell}, A.~M. 2018, J QUANT SPECTROSC RA, 217, 86

\bibitem[{{Wordsworth} \& {Pierrehumbert}(2014)}]{Wordsworth2014}
{Wordsworth}, R., \& {Pierrehumbert}, R. 2014, ASTROPHYS J, 785, L20

\bibitem[{Wordsworth {et~al.}(2011)Wordsworth, Forget, Selsis, Millour,
  Charnay, \& Madeleine}]{Wordsworth2011}
Wordsworth, R.~D., Forget, F., Selsis, F., {et~al.} 2011, ASTROPHYS J LETT,
  733, L48

\bibitem[{Zmuidzinas(2003)}]{Zmuidzinas2003}
Zmuidzinas, J. 2003, APPL OPTIC, 42, 4989.
\newblock \url{http://ao.osa.org/abstract.cfm?URI=ao-42-25-4989}

\end{thebibliography}


\begin{thebibliography}{}
\expandafter\ifx\csname natexlab\endcsname\relax\def\natexlab#1{#1}\fi
\providecommand{\url}[1]{\href{#1}{#1}}

\bibitem[{Fauchez {et~al.}(2017)Fauchez, Rossi, \& Stam}]{Fauchez2017}
Fauchez, T., Rossi, L., \& Stam, D.~M. 2017, ASTROPHYS J, 842, 41

\bibitem[{{Lincowski} {et~al.}(2018){Lincowski}, {Meadows}, {Crisp},
  {Robinson}, {Luger}, {Lustig-Yaeger}, \& {Arney}}]{Lincowski2018}
{Lincowski}, A.~P., {Meadows}, V.~S., {Crisp}, D., {et~al.} 2018, ASTROPHYS J,
  867, 76

\bibitem[{Lustig-Yaeger {et~al.}(2019)Lustig-Yaeger, Meadows, \&
  Lincowski}]{Lustig-Yaeger2019}
Lustig-Yaeger, J., Meadows, V.~S., \& Lincowski, A.~P. 2019, ASTRON J,
  \textbf{158}, 27.
\newblock \url{https://doi.org/10.3847%2F1538-3881%2Fab21e0}

\bibitem[{Misra {et~al.}(2014)Misra, Meadows, Claire, \& Crisp}]{Misra2014}
Misra, A., Meadows, V., Claire, M., \& Crisp, D. 2014, Astrobiology, 14, 67

\bibitem[{{Pall{\'e}} {et~al.}(2009){Pall{\'e}}, {Zapatero Osorio}, {Barrena},
  {Monta{\~n}{\'e}s-Rodr{\'{\i}}guez}, \& {Mart{\'{\i}}n}}]{Palle2009}
{Pall{\'e}}, E., {Zapatero Osorio}, M.~R., {Barrena}, R.,
  {Monta{\~n}{\'e}s-Rodr{\'{\i}}guez}, P., \& {Mart{\'{\i}}n}, E.~L. 2009,
  Nature, 459, 814

\bibitem[{Rinsland {et~al.}(1989)Rinsland, Zander, Namkung, Farmer, \&
  Norton}]{Rinsland1989}
Rinsland, C.~P., Zander, R., Namkung, J.~S., Farmer, C.~B., \& Norton, R.~H.
  1989, J GEOPHYS RES-ATMOS, 94, 16303

\bibitem[{{Schwieterman} {et~al.}(2016){Schwieterman}, {Meadows},
  {Domagal-Goldman}, {Deming}, {Arney}, {Luger}, {Harman}, {Misra}, \&
  {Barnes}}]{Schwieterman2016}
{Schwieterman}, E.~W., {Meadows}, V.~S., {Domagal-Goldman}, S.~D., {et~al.}
  2016, ASTROPHYS J, 819, L13

\bibitem[{Snellen {et~al.}(2013)Snellen, de~Kok, le~Poole, Brogi, \&
  Birkby}]{Snellen2013}
Snellen, I. A.~G., de~Kok, R.~J., le~Poole, R., Brogi, M., \& Birkby, J. 2013,
  ASTROPHYS J, 764, 182.
\newblock \url{https://doi.org/10.1088%2F0004-637x%2F764%2F2%2F182}

\end{thebibliography}

\paragraph*{Data availability}
The data that support the plots within this paper and other findings of this study are available from the corresponding author upon reasonable request.

\paragraph*{Code availability}
Atmos \citep{Arney2016} is available upon request from Giada Arney (giada.n.arney@nasa.gov); 
          LMD-G \citep{Wordsworth2011} is available upon request from Martin Turbet (martin.turbet@lmd.jussieu.fr);
          PSG \citep{Villanueva2018} is available on \url{https://psg.gsfc.nasa.gov/}.
\paragraph*{Acknowledgements}

T. Fauchez, G. Villanueva, G. Arney, R. Kopparapu, A. Mandell and S. Domagal-Goldman acknowledge support from GSFC Sellers Exoplanet Environments Collaboration (SEEC), which is funded in part by the NASA Planetary Science Divisions Internal Scientist Funding Model.\\ 
This project has received funding from the European Union’s Horizon 2020 research and innovation program under the Marie Sklodowska-Curie Grant Agreement No. 832738/ESCAPE.\\
This work was also supported by the NASA Astrobiology Institute Alternative Earths team under Cooperative Agreement Number NNA15BB03A and the NExSS Virtual Planetary Laboratory under NASA grant number 80NSSC18K0829. E.W.S. is additionally grateful for support from the NASA Postdoctoral Program, administered by the Universities Space Research Association. We thank Ha Tran for useful discussions related to O$_2$-X CIAs. Finally, we would like to thank the two anonymous referees for comments that greatly improved our manuscript.

\paragraph*{Author contributions}
T.J.F. led the photochemistry and transmission spectroscopy simulations. G.L.V, E.W.S and M.T. derived  parameterizations of the O$_2$-N$_2$ and O$_2$-CO$_2$ CIAs bands. T.J.F and G.A. wrote most of the manuscript.
Every author contributed to the discussions and to the writing of the manuscript.
\paragraph*{Reviewer information}
The authors declare no competing financial interests. Readers are welcome to comment on the online version of the paper.

\paragraph*{Competing interest}
The authors declare no competing financial interests.

\section*{Supplementary materials}

\paragraph{Overview of previous works on O$_2$ spectral features for exoplanet's studies.} Because O$_2$ is one of the most detectable and robust indicators of global biological activity, concepts for telescopes that would attempt to search for life on exoplanets all include the ability to detect O$_2$ or its photochemical byproduct, O$_3$. O$_2$ absorbs at several wavelengths in the visible (VIS) at 0.63, 0.69 and 0.76 $\mu m$ and near-infrared (NIR) at 1.27 $\mu m$. The O$_2$ A--band at 0.76 $\mu m$ has often been considered the most viable  spectral feature for oxygen detection in transmission \citep{Snellen2013} and reflectance spectra \citep{Fauchez2017}. \cite{Snellen2013} showed that it could be possible to detect the O$_2$ A--band in the atmosphere of an Earth twin with the future Extremely Large Telescopes (ELTs). However large unknowns remain to disentangle the exoplanet O$_2$ signal the from the telluric O$_2$. Meanwhile, \cite{Palle2009} showed that O$_2$--O$_2$ collision induced absorption (CIA) features at 1.06 and 1.27 $\mu m$ were present in Earth's transmission spectrum during lunar eclipse and produce more absorption than the O$_2$ A--band monomer feature.  CIA features are produced through inelastic collisions in a gas.  In the case of the O$_2$--O$_2$ CIA features, the two O$_2$ molecules interact forming transient multipole-induced dipoles producing broad spectral features distinct from the individual underlying O$_2$ molecule. \cite{Misra2014} showed that the 1.06 and 1.27 $\mu m$ O$_2$--O$_2$ CIA features may be detectable (for SNR~$>$~3) with the James Webb Space Telescope (JWST) for an Earth analogue orbiting an M5V star at a distance of 5~pc. \cite{Schwieterman2016} proposed that the 1.06 and 1.27 $\mu m$ transit features could be used to identify the high O$_2$ partial pressures predicted to be associated with abiotic O$_2$ atmospheres, which should be significantly higher than for the modern Earth case.   More recently, \cite{Lustig-Yaeger2019} have shown that the 1.06 and 1.27 $\mu m$ O$_2$ CIA features could be detectable with JWST at a SNR of 5 in just few transits for the TRAPPIST-1 planets with O$_2$ desiccated and dense (10 and 100~bars) atmospheres. \\

\makeatletter
\renewcommand{\fnum@figure}{Supplementary   \figurename~1}
\makeatother

\begin{figure}[H]
\centering
\resizebox{12cm}{!}{\includegraphics{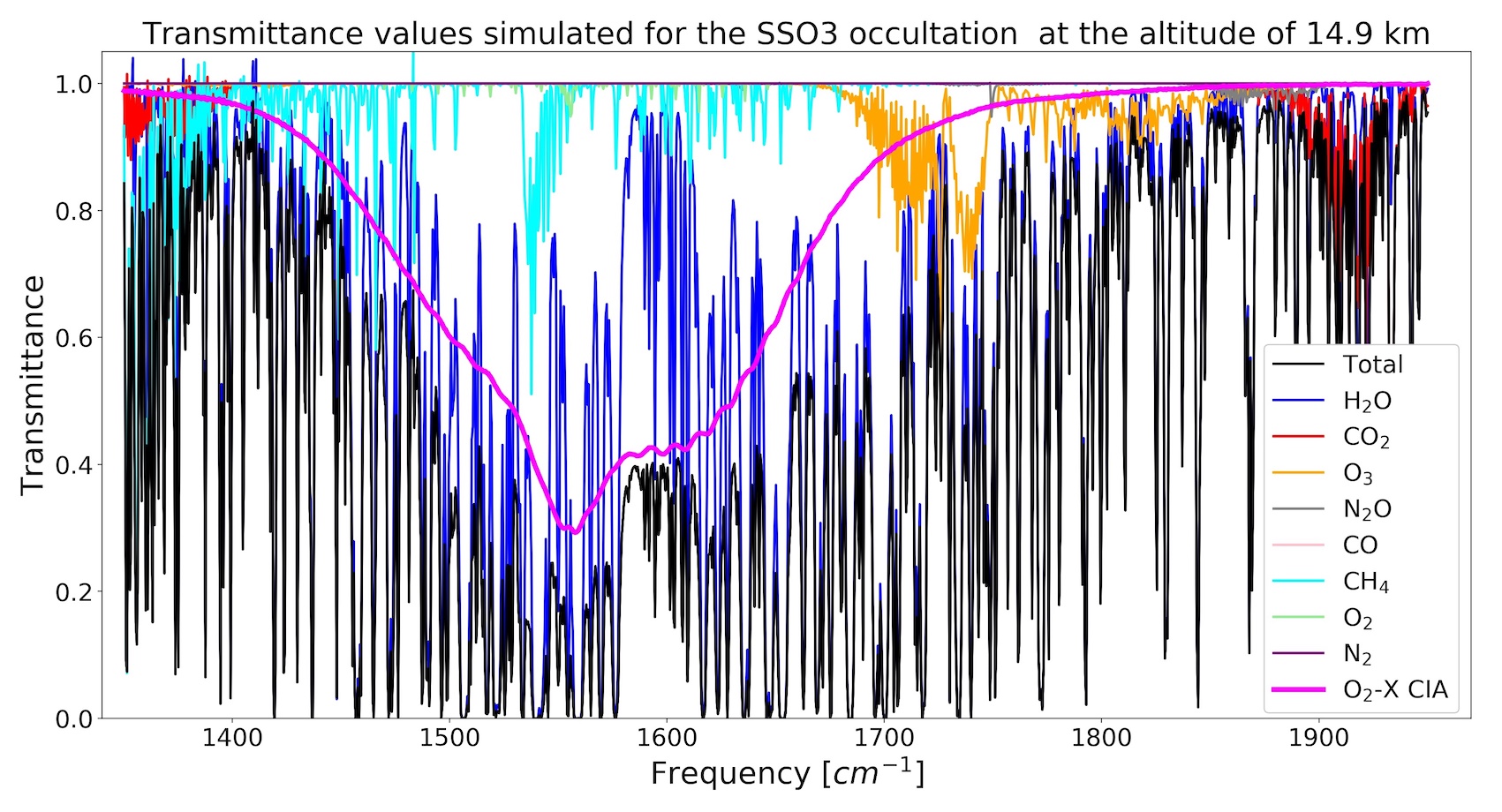}}
\caption{Simulation with PSG of the SSO3 occultation observed by \cite{Rinsland1989} April 30, 1985 at an altitude of 14.9~km over the latitude 32.3$^\circ$N and longitude 290.6$^\circ$W. Simulation of the $6.4\ \mu m$ O$_2$-X CIA is in very good agreement with observation data from \citep{Rinsland1989}.}
\label{fig:Rinsland}
\end{figure}

Supplementary Figure 1 shows the simulation of the Sun occultation SSO3 observed by \cite{Rinsland1989} (figure 3) on April 30, 1985, at an altitude of 14.9~km (13~km over the Himalayas). Our simulation and the observation data are in very good agreement, showing the validity of our O$_2$-X CIA parameterization at 6.4 $\mu m$.\\

Supplementary Figure 2 shows the terminator H$_2$O and O$_2$ atmospheric profiles with a modern Earth-like atmosphere composition for TRAPPIST-1 planets in the habitable zone, namely 1e, 1f and 1g  (top panel) and their transmission spectrum with clouds included (bottom panel). Boundary conditions for the photochemistry are those described in \cite{Lincowski2018} Table 8. We can see that the terminator region is very dry, with volume mixing ratios near the surface reaching the maximum value $10^{-3}$ for TRAPPIST-1e decreasing down to $10^{-6}$ for 1g. However, as we can see in the spectra this region of maximum H$_2$O concentration is below the continuum level because of clouds  and atmospheric refraction. Note that above $\sim60$~km H$_2$O is strongly photodissociated. As a results, the contribution of H$_2$O at $6.4\ \mu m$ is very largely dominated by the O$_2$-X CIA and this domination increase for dryer planets TRAPPIST-1f and 1g.\\

\makeatletter
\renewcommand{\fnum@figure}{Supplementary  \figurename~2}
\makeatother

\begin{figure}[H]
\centering
\resizebox{13cm}{!}{\includegraphics{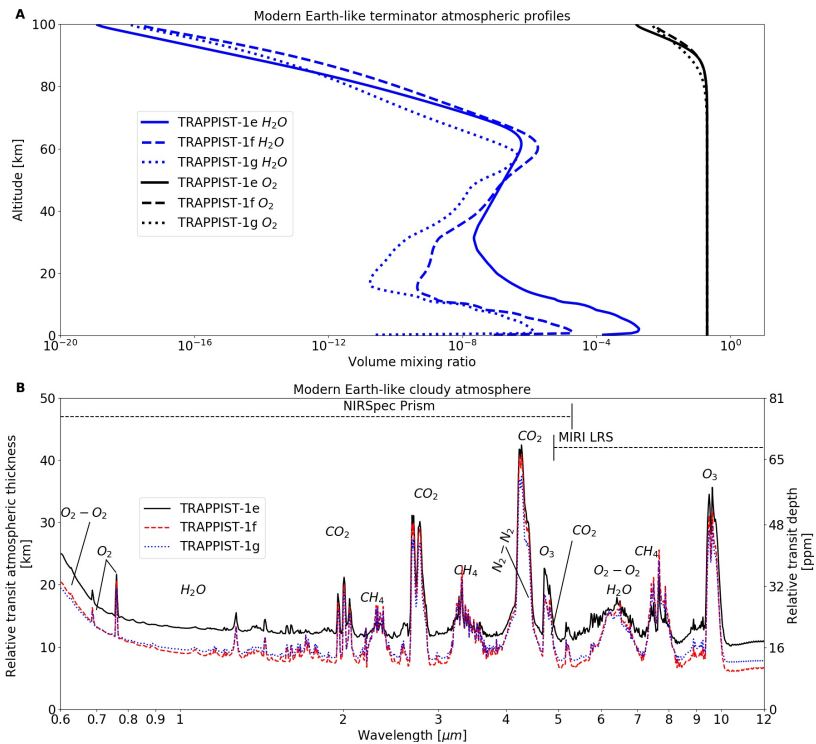}}
\caption{Panel A: H$_2$O and O$_2$ atmospheric profiles at the terminator of TRAPPIST-1e, 1f and 1g planets in the habitable zone with a modern Earth-like atmosphere. Panel B: Corresponding transmission spectra for the three planets. We can see that H$_2$O volume mixing ratio is tiny by comparison to O$_2$ and that the wetter region near the surface is below the continuum level of the spectra because of the atmospheric refraction and/or clouds. O$_2$-X largely dominates over H$_2$O in the $6.4\ \mu m$ region.}
\label{fig:profiles_spectra}
\end{figure}

Supplementary Figure 3 is similar to Fig. 2 but for the 22 bar O$_2$ desiccated and isothermal atmospheres presented in Table \ref{tab:transitDessic}. We can see that the O$_2$-O$_2$ $1.27\ \mu m$ CIA and O$_2$-X $6.4\ \mu m$ CIA features require significantly fewer transits than the O$_2$ A--band monomer band and would be detectable at up to about 25~pc (except for the coldest isothermal atmosphere beyond 20~pc). Note that in the case of a desiccated, O$_2$-rich planet with aerosols, the $6.4\ \mu m$ band would require significantly fewer transits than the $1.27\ \mu m$  for the same reasons as for the habitable case.\\

\makeatletter
\renewcommand{\fnum@figure}{Supplementary  \figurename~3}
\makeatother

\begin{figure}[H]
\centering
\resizebox{12cm}{!}{\includegraphics{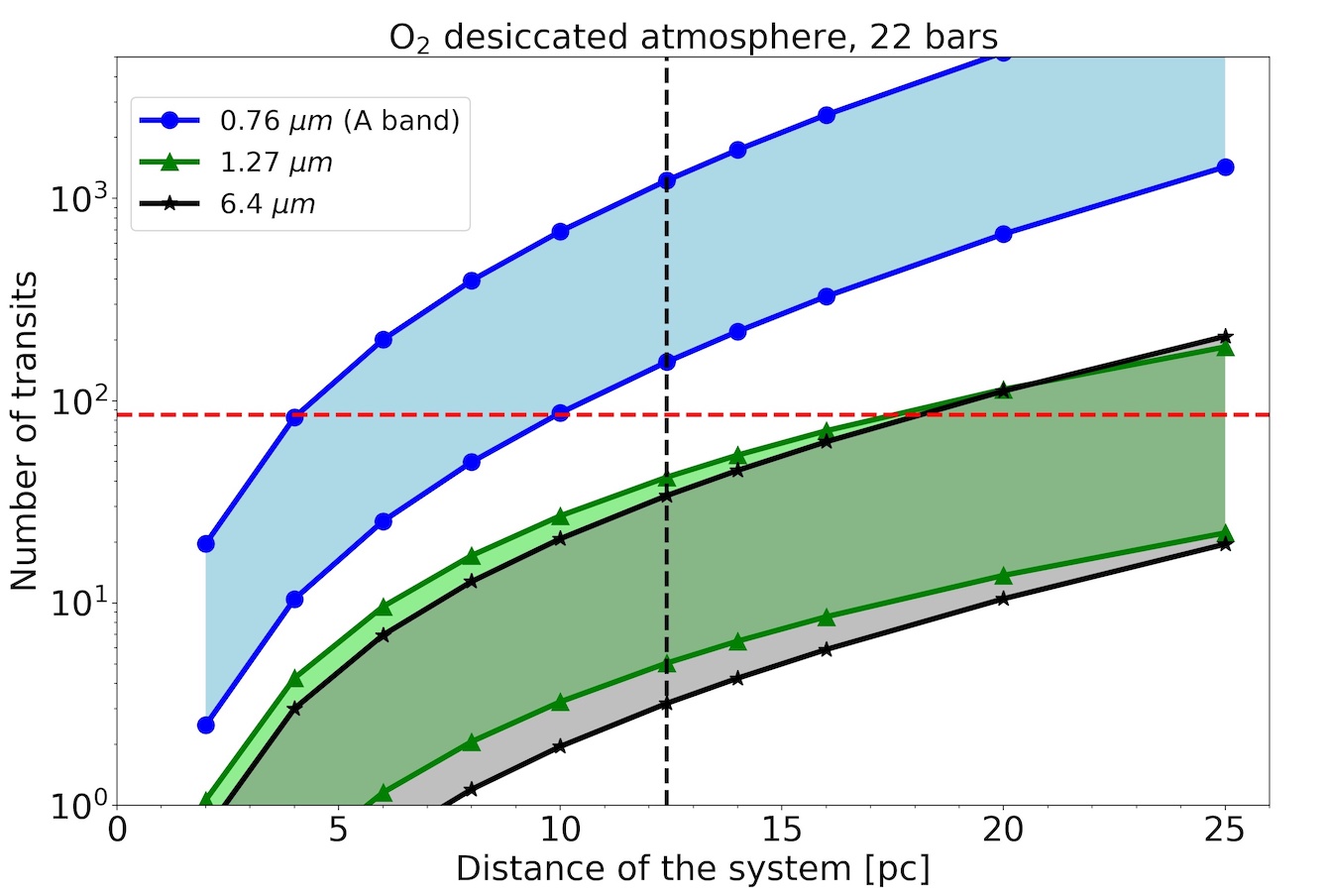}}
\caption{Number of TRAPPIST-1e transits needed for a $5\ \sigma$ detection of the O$_2$ A--band (R=100), the O$_2$-O$_2$ CIA at 1.27 $\mu m$ (R=20) and the O$_2$-X CIA at 6.4 $\mu m$ (R=10) with JWST for the TRAPPIST-1 system moved from 2 to 25~pc away from the Sun. The atmosphere is exclusively composed of O$_2$ with surface pressure of 22~bars. For each wavelength the shaded area correspond to various isothermal profiles from 600~K (lowest line) to 200~K (highest line).  Resolving power (R) has been optimized for each band to maximize the SNR. The horizontal dashed red line corresponds to the number of times TRAPPIST-1e will transits during JWST 5.5 years life time (85 transits). The vertical dashed black line denotes the distance of the TRAPPIST-1 system with respect to the Sun.  The O$_2$-O$_2$ CIA at 1.27 $\mu m$ and the O$_2$-X CIA at 6.4 $\mu m$ are detectable up to 25~pc away, except for the coldest atmospheres beyond 20~pc.}
\label{fig:dessicTransits}
\end{figure}

Supplementary  Table \ref{tab:transitDessic} presents the relative transit depth, 1 transit SNR and number of transits for 3 and 5 $\sigma$ detections for TRAPPIST-1e assuming 1 and  22~bar desiccated atmosphere on TRAPPIST-1e. 22 bars is based on a conservative estimate of O$_2$ retention by  \cite{Lincowski2018}.  We can see that the difference in transit depth between the 1 and 22 bar cases increase with temperature (because the refraction limit is at higher pressures) and that the strength of O$_2$ A--band is relatively insensitive to pressure. The O$_2$-X CIA feature at 6.4 $\mu m$ requires fewer transits to achieve 3 or 5 $\sigma$ detection and is therefore the most promising indicator of a massive O$_2$ desiccated atmosphere potentially observable with JWST.\\

\makeatletter
\renewcommand{\fnum@table}{Supplementary  \tablename~1}
\makeatother
\begin{table}[H]
\centering
\caption{Relative transit depth (ppm), signal-to-noise ratio  for 1 transit (SNR-1) and number of transits to achieve a $5\ \sigma$ and $3\ \sigma$ detection of O$_2$ assuming O$_2$ desiccated and isothermal atmospheres on TRAPPIST-1e \cite{Lincowski2018}. The numbers at the left of the "--" mark are for the 1~bar atmosphere while the numbers at the right are for the 22~bar atmosphere. (-) represent the cases for which more than 100 integrated transits are needed. For each feature, the wavelength and  resolving power (R) are mentioned.} \label{tab:transitDessic}
\begin{tabular}{c c c  c c}
\hline
\hline
Feature & A--band & O$_2$-O$_2$ & O$_2$-O$_2$ & O$_2$-X   \\
Wavelength [$\mu m$] & 0.76 & 1.06 & 1.27 & 6.4   \\
R & 100 & 40 & 20 & 10   \\
\hline
Temperature & \multicolumn{4}{c}{200~K}\\
Depth [ppm] & 44--44 & 38--37 & 42--41 & 67--66 \\
SNR-1  &  0.25--0.25 & 0.66--0.65 &  1.16--1.14 & 1.33--1.31\\
N transits  ($5\sigma$) & (-) & 57--59 & 19--19 &  14--15 \\
N transits  ($3\sigma$) & (-) & 21--21 &  7--7 &  5--5\\
\hline
Temperature & \multicolumn{4}{c}{300~K}\\
Depth [ppm] & 68--68  & 52--62 &  57--71 & 88--107 \\
SNR-1  & 0.39--0.39  &  0.90--1.07 & 1.59--1.98  & 1.75--2.13\\
N transits  ($5\sigma$) & (-) & 31--22 & 10--7 &  8--6 \\
N transits  ($3\sigma$) & 59--59 & 11--8 & 4--3  & 3--2 \\
\hline
Temperature & \multicolumn{4}{c}{400~K}\\
Depth [ppm] & 91--99 & 63--88  &  71--107 & 110--162\\
SNR-1  &  0.52--0.57 & 1.10--1.54 &  1.97--2.97 & 2.18--3.22 \\
N transits  ($5\sigma$) & 93--77 & 21--11 & 6--3 & 5--2  \\
N transits  ($3\sigma$) & 33--28 & 7--4 & 2--1  & 2--1 \\
\hline
Temperature & \multicolumn{4}{c}{500~K}\\
Depth [ppm] & 114--127 & 74--108 & 83--132 & 129--197\\
SNR-1  & 0.65--0.73  & 1.28--1.88 & 2.31--3.65  & 2.56--3.91\\
N transits  ($5\sigma$) & 59--47 & 15--7 & 5--2  &  4--2 \\
N transits  ($3\sigma$) & 21--17 & 6--3 &  2--1  &  1--1\\
\hline
Temperature & \multicolumn{4}{c}{600~K}\\
Depth [ppm] & 137--156 & 83--136 &  95--174 & 147--264 \\
SNR-1  &  0.79--0.89 & 1.44--2.37 &  2.63--4.83 & 2.93--5.23\\
N transits  ($5\sigma$) & 40--32 & 12--5 & 4--1 &  3--1 \\
N transits  ($3\sigma$) & 14--12  & 4--2 &  1 & 1--1 \\
\hline
\hline
\end{tabular}
\end{table}

\end{document}